\documentclass[12pt]{article}
\newcommand{\be}{\begin{equation}}
\newcommand{\ee}{\end{equation}}
\newcommand{\bea}{\begin{eqnarray}}
\newcommand{\eea}{\end{eqnarray}}
\newcommand{\sptwo}{1.4}
\newcommand{\doublespace}{\edef\baselinestretch{\sptwo}\Large\normalsize}
\newcommand{\newsection}[1]{
\section{#1}
\setcounter{equation}{0}}

\renewcommand{\theequation}{\thesection.\arabic{equation}}
\newcounter{newapp}
\renewcommand{\thenewapp}{\Alph{newapp}}
\begin{document}
\begin{center}
{\large\bf Brane Dynamics \\
From \\
Non-Linear Realizations}
\end{center}
~\\
\begin{center}
{\bf T.E. Clark\footnote{e-mail address: clark@physics.purdue.edu} and
Muneto
Nitta\footnote{e-mail address: nitta@physics.purdue.edu}}\\
{\it Department of Physics\\
Purdue University\\
West Lafayette, IN 47907-1396}\\
~\\
And\\
~\\
{\bf T. ter Veldhuis\footnote{e-mail address: veldhuis@physics.umn.edu}}\\

{\it Department of Physics\\
University of Minnesota\\
Minneapolis, MN 55455}
~\\
~\\
\end{center}
\begin{center}
{\bf Abstract}
\end{center}
The action for a non-BPS p=2 brane embedded in a flat N=1, D=4 target 
superspace
is obtained through the method of nonlinear realizations of the associated
super-Poincar\'{e} symmetries.  The brane excitation modes correspond to the
Nambu-Goldstone degrees of freedom resulting from the broken space
translational symmetry and the target space supersymmetries.  The action for
this p=2 brane is found to be an invariant synthesis of the Akulov-Volkov
and Nambu-Goto actions.  The dual D2-brane Born-Infeld action is derived.
The invariant coupling of matter fields localized on the brane to the
Nambu-Goldstone modes is also obtained.
~\\
~\\
\newpage
\doublespace

\newsection{\large Introduction}

A domain wall spontaneously  breaks the Poincar\'e invariance of the target
space down to the symmetry group of the world volume subspace of the wall,
which includes a lower dimensional Poincar\'e symmetry.  The long wavelength
oscillation modes of the domain wall are described by the Nambu-Goldstone
bosons associated with the collective coordinate translations transverse to
the wall.  Indeed, the Nambu-Goto action governing the zero mode fields'
dynamics is easily obtained in a model independent way by nonlinearly
realizing the broken symmetries on the Nambu-Goldstone fields
\cite{Coleman:sm,Volkov73}.  In the case
of a two dimensional domain wall (or p=2-brane) embedded in three
dimensional space, the D=3 Poincar\'e generators, $p^m$ for space-time
translations and $M^m = \frac{1}{2} \epsilon^{mnr} M_{nr}$ for Lorentz
rotations, form an unbroken subgroup $H=ISO(1, 2)$ of the D=4 Poincar\'e
group $G=ISO(1, 3)$.  The broken generators are the D=4 translation generator
transverse to the wall which is a D=3 Lorentz scalar, denoted $Z$, and
the three broken D=4 Lorentz rotations which form a D=3 Lorentz
vector, denoted $K^m$.  The D=4 Poincar\'e group can
be realized by group elements acting on the coset $ISO(1, 3)/SO(1, 2)$
element $\Omega$ formed from the $p^m ,~Z ,~K^m$ charges
\be
\Omega \equiv e^{i x^m p_m } e^{i \phi(x) Z} e^{iv^m(x) K_m} ,
\ee
where the world volume D=3 space-time coordinates of the 2-brane wall in the
static gauge are $x^m $, while $\phi (x)$ and $v^m (x)$ are the collective
coordinate Nambu-Goldstone bosons associated with the broken D=4 Poincar\'e
symmetries corresponding to the excitation modes of the 2-brane.  The D=4
Poincar\'e group transformations are realized by left multiplication by
group elements $g$
\be
g\Omega = \Omega^\prime h ,
\label{transform}
\ee
where the new coset element $\Omega^\prime$ has the form
\be
\Omega^\prime = e^{i x^{\prime m} p_m } e^{i \phi^\prime ( x^\prime ) Z}
e^{i v^{\prime m} ( x^\prime ) K_m},
\ee
and yields the transformation law for the coordinates and fields and
\be
h = e^{i \beta^m  (g, v ) M_m }
\ee
allows $\Omega^\prime$ to be written as a coset element.  The set of charges
$\{ p^m , M^{m} \}$ generate the vacuum stability group $H$ of the
system and are linearly represented.  For the general set of infinitesimal
transformations $g \in G$,
\be
g= e^{i [ a^m p_m + zZ + b^m K_m + \alpha^m M_m]},
\ee
the D=4 Poincar\'e algebra, written in D=3 Lorentz group form
\begin{center}
\begin{tabular}{ll}
$[p^m , p^n ] = 0$ & $[M^m ,M^n ] = -i\epsilon^{mnr} M_r$ \\
$[p^m , Z ] = 0$ & $[M^m ,K^n ] = -i\epsilon^{mnr} K_r $ \\
& $[K^m , K^n ] = +i\epsilon^{mnr} M_r$ \\
& \\
$[M^{m}, p^n ] = -i\epsilon^{mnr} p_r\;\;\;\;\;$ & $[K^m , p^n ] = 
+i\eta^{mn}
Z$ \\
$[M^{m} , Z ] = 0$ & $[K^m , Z ] = + i p^m$,
\end{tabular}
\end{center}
\be
~
\label{Poincarealg0}
\ee
can be exploited to find the space-time coordinate variations and field
transformations
\bea
x^{\prime~m} &=& x^m + a^m -\phi b^m + \epsilon^{mnr} \alpha_n x_r \cr
\Delta \phi &=& z - b_m x^m \cr
\Delta v^m &=& \frac{ \sqrt{v^2} }{\tanh{\sqrt{v^2}}} b^m + \left(1-\frac{
\sqrt{v^2} }{\tanh{\sqrt{v^2}}}\right) \frac{b_n v^n v^m}{v^2} +
\epsilon^{mnr} \alpha_n v_r .
\eea
Here the field transformations are total variations so that $\Delta  \phi
(x) = \phi^\prime (x^\prime ) - \phi (x)$, and likewise for $v^m$.

Constructing the Maurer-Cartan world volume one-forms,
\be
\Omega^{-1} d\Omega \equiv i\left[ \omega^a p_a + \omega_{Z} Z +
\omega_{K}^{a} K_a + \omega_{M}^{a} M_a \right] ,
\label{mc1forms1}
\ee
defines the dreibein, $e_m^{~a}$, which relates the covariant world volume
coordinate differentials $\omega^a$ to the world volume coordinate
differentials $dx^m$, so that $\omega^a = dx^m e_m^{~a}$, the covariant
derivatives of the fields, $\omega_Z \equiv \omega^a \nabla_a \phi$ and
$\omega_K^b \equiv \omega^a \nabla_a v^b$ , and the spin connection
$\omega_M^b \equiv \omega^a \Gamma_a^{~b}$.  Once again utilizing the D=4
Poincar\'e algebra,
the dreibein is found to be
\be
e_m^{~a} = \delta_m^{~a} +\left[ \cosh{\sqrt{v^2}} -1\right] \frac{v_m
v^a}{v^2} + \partial_m \phi v^a \frac{\sinh{\sqrt{v^2}}}{\sqrt{v^2}} ,
\label{e1}
\ee
while the $\phi$-field covariant derivative is
\be
\omega_Z = \omega^a \nabla_a \phi = dx^m e_m^{~a} \nabla_a \phi = dx^m
\cosh{\sqrt{v^2}} \left[ \partial_m \phi + v_m
\frac{\tanh{\sqrt{v^2}}}{\sqrt{v^2}} \right] .
\label{phiderivative1}
\ee
The one-form transformation laws follow from equation (\ref{transform})
\be
\left( \Omega^{-1} d \Omega \right)^\prime = h \left( \Omega^{-1} d \Omega
\right) h^{-1} + h d h^{-1},
\ee
and are homogeneous except for the case of the broken D=4 Lorentz rotations
generated by $K^n$, in which case
\be
h = e^{-\frac{i}{2}
\frac{\tanh{\frac{1}{2}\sqrt{v^2}}}{\frac{1}{2}\sqrt{v^2}}
b_m v_r \epsilon^{mrn} M_n },
\ee
implying that $\omega_M^{~m}$ transforms with an additional inhomogeneous
term as required of a connection one-form.

Given these building blocks and their transformation laws, the low energy
$G$-invariant action, $\Gamma$, is obtained in leading order in the domain
wall (brane) tension $\sigma$
\be
\Gamma = -\sigma \int d^3 x \det{e} ,
\ee
with the determinant of $e$ determined to be
\be
\det{e} = \cosh{\sqrt{v^2}}\left[ 1 +\partial_n \phi v^n
\frac{\tanh{\sqrt{v^2}}}{\sqrt{v^2}}\right].
\label{dete}
\ee
Since the dreibein depends only on $v^m$ and not its derivatives, it's
Euler-Lagrange equation of motion can be used to eliminate $v^m$ in terms of
$\phi$.  This is just the \lq\lq inverse Higgs mechanism\rq\rq
\cite{Ivanov:1975zq}, equivalently obtained by setting the $\phi$ covariant
derivative, equation (\ref{phiderivative1}), to zero: $\nabla_a \phi =0$ and
hence
\be
v^m \frac{\tanh{\sqrt{v^2}}}{\sqrt{v^2}} = -\partial^m \phi .
\ee
Substituting this into the dreibein, it has the form
\bea
e_m^{~a} &=& \delta_m^{~a} + \left(
\frac{1-\cosh{\sqrt{v^2}}}{\cosh{\sqrt{v^2}}}\right) \frac{v_m v^a}{v^2} \cr
&=& \delta_m^{~a} -\left( 1-\sqrt{1-(\partial\phi)^2} \right)
\frac{\partial_m \phi \partial^a \phi}{(\partial \phi)^2} .
\eea
The determinant of $e$ simplifies to become
\be
\det{e} = \frac{1}{\cosh{\sqrt{v^2}}} = \sqrt{1-\partial_m \phi \partial^m
\phi } ,
\ee
and the Nambu-Goto action \cite{Dirac:1962iy,Nambu:1974zg,Goto:ce}
for a p=2 brane embedded in D=4 space-time (in the
static gauge) is obtained
\be
\Gamma = -\sigma \int d^3 x \sqrt{1-\partial_m \phi \partial^m \phi } .
\label{NGaction1}
\ee

Alternatively, the $\phi$ and $v^m$ fields can be kept as independent
degrees of freedom.  The action is given in terms of equation (\ref{dete}).
The $\phi$ equation of motion, $\delta \Gamma / \delta \phi =0$, can be
expressed as the D=3 Bianchi identity, $\partial_m F^m =0$, for the field
strength vector
\be
F^m \equiv v^m \frac{\sinh{\sqrt{v^2}}}{\sqrt{v^2}} .
\ee
Substituting this into equation (\ref{dete}) yields
$\det{e} = \sqrt{1+F^2} +\partial_m \phi F^m$.  Integrating
the second term by parts and using $\partial_m F^m =0$ implies
duality of the Nambu-Goto p=2 brane action to the Born-Infeld action
\cite{Born:gh} for a
D2-brane
\be
\Gamma = -\sigma \int d^3 x \det{e} = -\sigma \int d^3 x \sqrt{1 + F^2} .
\ee

A slightly generalized approach can be applied to the above coset method as
described in \cite{West:2000hr}. The brane world volume is parameterized by
the D=3 vector $\xi^m$ and the brane's generalized coordinates are maps of
this D=3 parameter space into the D=4 target manifold:
$x^\mu (\xi) = ( x^m(\xi), \phi (\xi))$.  
The exterior derivative is given by
$d=d\xi^m \partial / \partial \xi^m $ and it is world volume
reparameterization invariant.  The Maurer-Cartan one-forms, equation
(\ref{mc1forms1}), are also reparameterization invariant since they depend
on the exterior derivative.  From this point of view the covariant
differential one-forms, $\omega^a$, define the dreibein as $\omega^a \equiv
d\xi^m e_m^{~a}$ where now
\be
e_m^{~a} = \frac{\partial x^a}{\partial \xi^m} + \left[ \cosh{\sqrt{v^2}}
-1\right] \frac{v_b \frac{\partial x^b}{\partial \xi^m} v^a}{v^2} +
\frac{\partial \phi}{\partial \xi^m} v^a
\frac{\sinh{\sqrt{v^2}}}{\sqrt{v^2}} .
\ee
Similarly the covariant differential one-form $\omega_Z$ of the $\phi (\xi)$
coordinate is given by its covariant derivative
\be
\omega_Z = \omega^a \nabla_a \phi = d \xi^m \cosh{\sqrt{v^2}} \left[
\frac{\partial \phi}{\partial \xi^m} + \frac{\partial x^a}{\partial \xi^m}
v_a \frac{\tanh{\sqrt{v^2}}}{\sqrt{v^2}} \right] .
\ee

The Maurer-Cartan one-form can be calculated sequentially as
\bea
\Omega^{-1} d \Omega &=& e^{-iv^n K_n} \left( d + i dx^\mu P_\mu \right)
e^{+iv^n K_n} \cr
&=& idx^\mu \Lambda_\mu^{~\nu} (v) P_\nu + e^{-iv^n K_n} d e^{+iv^n K_n} ,
\eea
where $\Lambda_\mu^{~\nu} (v)$ is a broken 
D=4 Lorentz transformation determined by $v^n$, and $P_\mu = (p_m, Z)$.   
Thus it is seen that the one-forms
$\omega^\mu \equiv (\omega^a , \omega_Z)$ are related
to $dx^\mu$ by a Lorentz transformation
\be
\omega^\mu = dx^\nu \Lambda_\nu^{~\mu} (v) .
\ee
From the invariance of the D=4 Minkowski metric $\eta_{\mu\nu}$ under D=4
Lorentz transformations, it follows that
\bea
\omega^\mu \eta_{\mu\nu} \omega^\nu  &=& dx^\mu \Lambda_\mu^{~\rho}
\eta_{\rho\sigma} \Lambda_\nu^{~\sigma} dx^\nu \cr
&=& dx^\mu \eta_{\mu\nu} dx^\nu .
\eea
As before, the inverse Higgs mechanism may be applied, $\omega_Z =0$,
yielding
\bea
\omega^\mu \eta_{\mu\nu} \omega^\nu  &=& d\xi^m e_m^{~a} \eta_{ab} e_n^{~b}
d\xi^n \cr
&=& d\xi^m \frac{\partial x^\mu}{\partial \xi^m} \eta_{\mu\nu}
\frac{\partial x^\nu}{\partial \xi^n} d\xi^n .
\eea
This is just the $G$-invariant interval, hence the world volume
reparameterization invariant and $G$-invariant action is
\be
\Gamma = -\sigma \int d^3 \xi \det{e} = -\sigma \int d^3 \xi \sqrt{
\det{\left(\frac{\partial x^\mu}{\partial \xi^m} \eta_{\mu\nu}
\frac{\partial x^\nu}{\partial \xi^n} \right)}}.
\ee
Thus, the general form of the Nambu-Goto action for a p=2 brane is secured.
The reparameterization invariance may be used to fix the static gauge: $x^m
= \xi^m$ and $\phi = \phi (x)$, in which case the action reduces to that of
equation (\ref{NGaction1}).  The remainder of the paper is in the static
gauge.

The above considerations can be generalized to apply in a supersymmetric
context by embedding a topological defect in superspace. Apart from 
Goldstone
bosons associated with spontaneously
broken translational invariances, there are in this case additional 
fermionic
long wavelength oscillations. These
Goldstinos reflect collective Grassmann coordinates
which are associated with spontaneously broken supersymmetries. Additional
massless world volume degrees of freedom may be required to complete
multiplets of the unbroken supersymmetries.  Topological defects which
spontaneously break down target space super-Poincar\'e
invariance to a lower dimensional
super-Poincar\'e symmetry were considered by \cite{Hughes:fa,Hughes:dn}.
If the spatial extension of a defect in directions of broken
translational invariance is small
compared to the wavelength of its
fluctuations, and if in addition some
supersymmetry remains unbroken, then such a defect is a
super p-brane \cite{Duff:1996zn,West:1998ey}. The world volume
theory on the defect inherits extended supersymmetry.
Part of this supersymmetry as well as central charges corresponding
to spontaneously broken translation generators of the target space are
nonlinearly realized \cite{Bagger:1994vj}.

The previous illustrative example dealt with a D=3 space-time
world volume of the p=2 brane
being embedded in a target D=4 space-time. Alternatively, the D=3
space-time world volume can be embedded in a target N=1, D=4 superspace;
this is the case of a non-BPS brane embedded into N=1, D=4 superspace,
the main topic of this paper. When
such a 2-brane domain wall is embedded into superspace, all supersymmetry is
spontaneously broken as well as the spatial translation symmetry.

If in contrast the defect is a BPS domain wall, then the supersymmetry
is only partially broken \cite{deAzcarraga:gm}. 
The tension saturates its lower bound, which is equal to the absolute
value of the central charge, and the domain wall is therefore stable. 
The world volume of the corresponding super 2-brane is N=1, D=3
superspace. In the thin wall limit its dynamics were studied 
using the method of
nonlinear realizations \cite{Ivanov:1999fw,Ivanov:2001gd} as well
as equivalently using the superembedding technique \cite{Sorokin:1999jx}.
The world volume theory
of BPS domain walls with finite width was also studied
\cite{Chibisov:1997rc,Sakamura:2002iw} through
an expansion
in modes about classical domain wall solutions.

BPS saturated domain walls provide an effective mechanism for the partial
breaking of supersymmetry and  may even be a necessary
ingredient in a more
fundamental brane world and M-theory description of nature. On the other
hand, a non-BPS domain wall can be stable and as such can provide a means to
completely break the supersymmetry. The lower dimensional manifold of the
domain wall will then also include the corresponding Goldstino modes besides
the broken translational symmetry Nambu-Goldstone boson mode.  These fields
correspond to the excitations of the brane in all possible target space
directions, the space direction orthogonal to the brane and in this case
the Grassmann coordinate directions $\theta_\alpha$
and $\bar\theta_{\dot\alpha}$ of N=1,
D=4 superspace when in the static gauge.  It is the purpose of this paper to
construct the effective action via the method of nonlinear realizations for
these low energy degrees of freedom.  In addition to the massless
Nambu-Goldstone fields of the 2-brane motion, there also can be light
matter field degrees of freedom
localized on the domain wall brane.  Their invariant interaction with the
Nambu-Goldstone fields is determined as well.

Section 2 analyzes the method of the nonlinear realization of N=1, D=4
super-Poincar\'{e} symmetries on the Nambu-Goldstone fields as coset manifold
coordinates.  The associated Maurer-Cartan one-forms are constructed
in section 3.  Included in these is the D=3 world volume dreibein which is
used to construct the N=1, D=4 super-Poincar\'{e} invariant action.
In section 4, the covariant derivatives of the Nambu-Goldstone fields, 
obtained
from the Maurer-Cartan one-forms, are shown to provide a means
to covariantly reduce the number of fields, through the "inverse Higgs
mechanism" \cite{Ivanov:1975zq}, to the minimal set of independent degrees 
of
freedom needed to describe the fluctuations of the 2-brane
in N=1, D=4 superspace.  This description of the brane dynamics is performed 
in
the partially covariant one-form basis which reveals the product
nature of the world volume dreibein and leads directly to the
invariant synthesis of the Nambu-Goto and Akulov-Volkov actions.  
Alternatively, exploiting the general form of the brane action
in terms of all fields, the duality between the Nambu-Goto-Akulov-Volkov 
action and the
D2-brane (non linearly realized) supersymmetric Born-Infeld action 
is derived.  Finally, in section 5,
the invariant action describing scalar and fermion matter fields localized 
on
the brane is constructed.  

The brane and matter field actions are the lowest
order terms in an expansion in powers of the domain wall thickness.
In this approximation the 2-brane is thin
relative to its fluctuation wavelength, but the amplitude of
the fluctuations may be large.
The covariant derivatives for Goldstone and matter fields 
determined in sections 4 and 5 form the building
blocks from which higher order terms in the expansion can be constructed
in order to obtain an action that describes 
large amplitude, shorter wavelength (but still larger than the
domain wall thickness) fluctuations as well.
Additional higher order terms have coefficients that parameterize
in the world volume theory indirect effects of massive modes which exist
in the underlying fundamental theory. Such massive modes generically
have masses proportional to the inverse of the domain wall width.
The remainder of the
introduction outlines the results derived in sections 2-5 of the body of the 
paper.

The Nambu-Goldstone modes' action is an invariant synthesis of the
Akulov-Volkov action \cite{Volkov:jx} and the Nambu-Goto action.  This
action consists of a product of the Akulov-Volkov lagrangian and a modified
Nambu-Goto lagrangian allowing for excitations of the non-BPS brane in the
Grassmann coordinate directions of the target superspace
\be
\Gamma = -\sigma \int d^3 x \det{e} = -\sigma \int d^3 x
\det{\hat{e}}\det{N} ,
\label{ngav-action}
\ee
where the 2-brane dreibein is given by a product of dreibeine $e_m^{~a} =
\hat{e}_m^{~b} N_b^{~a}$.  The Akulov-Volkov dreibein $\hat{e}_m^{~a}$ is
\be
\hat{e}_m^{~a} =A_m^{~a}= \delta_m^{~a} + i \partial_m \theta
\gamma^0\gamma^a \theta + i \partial_m \lambda \gamma^0 \gamma^a \lambda ,
\ee
with the Goldstino fields given by the D=3 Majorana spinors $\theta_i (x) $ 
and
$\lambda_i (x)$.  The Akulov-Volkov determinant term in the action has its
typical form
\be
\det{\hat{e}}= \det{\left[ \delta_m^{~a} + i \partial_m \theta
\gamma^0\gamma^a \theta + i \partial_m \lambda \gamma^0 \gamma^a \lambda
\right]}.
\label{AV}
\ee
The Nambu-Goto dreibein $N_a^{~b}$ is given by a supersymmetric extension of
equation (\ref{e1}) above
\be
N_a^{~b} = \delta_a^{~b} +\left[ \cosh{\sqrt{v^2}} -1\right] \frac{v_a
v^b}{v^2} + \left( {\hat{\cal D}}_a \phi +   {\hat{\cal D}}_a \theta
\gamma^0 \lambda - \theta \gamma^0 {\hat{\cal D}}_a \lambda \right)  v^b
\frac{\sinh{\sqrt{v^2}}}{\sqrt{v^2}} ,
\label{N}
\ee
where ${\hat{\cal D}}_a = \hat{e}_a^{-1 m}\partial_m$ is the Akulov-Volkov
partial covariant derivative.  The determinant of the Nambu-Goto dreibein is
found to be
\be
\det{N} = \cosh{\sqrt{v^2}} \left[ 1 + \left( {\hat{\cal D}}_a \phi +
{\hat{\cal D}}_a \theta \gamma^0 \lambda - \theta \gamma^0 {\hat{\cal D}}_a
\lambda \right)  v^a \frac{\tanh{\sqrt{v^2}}}{\sqrt{v^2}}\right].
\label{detN}
\ee

As with the dreibein, the Maurer-Cartan one-form associated with the central
charge $Z$ has a supersymmetric generalization to include motion in the
anticommuting directions
\be
\omega_Z = \cosh{\sqrt{v^2}}\left[ \left( d\phi +d\theta \gamma^0 \lambda 
-\theta
\gamma^0 d\lambda \right) + dx^m \hat{e}_m^{~a} v_a
\frac{\tanh{\sqrt{v^2}}}{\sqrt{v^2}} \right] .
\ee
Setting it to zero once again leads to the \lq\lq inverse Higgs
mechanism\rq\rq
\be
v_a \frac{\tanh{\sqrt{v^2}}}{\sqrt{v^2}} = - \left({\hat{\cal D}}_a \phi +
{\hat{\cal D}}_a \theta \gamma^0 \lambda - \theta \gamma^0 {\hat{\cal D}}_a
\lambda \right)  .
\ee
Thus the super Nambu-Goto determinant reduces to
\be
\det{N} = \frac{1}{\cosh{\sqrt{v^2}}} = \sqrt{1- \left({\hat{\cal D}}_a \phi
+   {\hat{\cal D}}_a \theta \gamma^0 \lambda - \theta \gamma^0 {\hat{\cal
D}}_a \lambda \right)^2} .
\label{NG}
\ee
Hence, the D=4 super-Poincar\'e invariant action is obtained from the
product of equations (\ref{AV}) and (\ref{NG}).

The domain wall world volume embedded in superspace is dual to the D2-brane
embedded in superspace \cite{Duff:1992hu,Aganagic:1996nn} as is expressed by 
the
above Nambu-Goto-Akulov-Volkov
action being dual to the supersymmetric Born-Infeld action.  Treating all
fields as independent degrees of freedom,
the $\phi$ field equation is obtained from equations (\ref{AV}) and
(\ref{detN}) above and takes the form of the D=3 Bianchi identity,
$\partial_m F^m =0$, with the field strength vector now given by
\be
F^m \equiv \det{\hat{e}}\frac{\sinh{\sqrt{v^2}}}{\sqrt{v^2}} v^a
\hat{e}_a^{-1 m} .
\ee
Substituting this into the determinant of the dreibein yields
\be
\det{\hat{e}}\det{N} = \det{\hat{e}}\cosh{\sqrt{v^2}} + F^n \left(\partial_n
\theta \gamma^0 \lambda - \theta \gamma^0 \partial_n \lambda \right) + F^n
\partial_n \phi .
\ee
The last term integrates to zero to obtain that the non-BPS p=2 brane
Nambu-Goto-Akulov-Volkov action is dual to the D2-brane supersymmetric
Born-Infeld action
\bea
\Gamma &=& -\sigma \int d^3 x \left(\sqrt{ \det{\left( \hat{g}_{mn} +
F_{mn}\right)}} +F^n [\partial_n \theta \gamma^0 \lambda -\theta \gamma^0
\partial_n \lambda ]\right) \cr
&=& -\sigma \int d^3 x \left(\sqrt{ \det{\hat{g}} + F^m \hat{g}_{mn} F^n }
+F^n [\partial_n \theta \gamma^0 \lambda -\theta \gamma^0 \partial_n \lambda
]\right),
\eea
where the Akulov-Volkov metric is given by $\hat{g}_{mn} =
\hat{e}_m^{~a}\eta_{ab}\hat{e}_n^{~b}$ and $F_{mn} = \epsilon_{mnr} F^r$.

A specific example of an underlying field theory realizing a stable non-BPS
domain wall can be constructed as a
generalized Wess-Zumino model in D=4 dimensions \cite{Chibisov:1997rc}. It
contains two chiral superfields $X$ and $\Phi$, with
superpotential $W=X(\mu^2 - \lambda\Phi^2)$ and canonical
K\"{a}hler potential. The tension $\sigma=\frac{8}{3} \mu^3/\sqrt{\lambda}$
of the domain wall that interpolates between
the two vacua $X=0,\;\Phi=\pm\sqrt{\mu^2/\lambda}$ does
not saturate the BPS bound $\left|2 \Delta W\right|=0$, yet
the wall is stable. The width of the wall is $1/\sqrt{\mu^2 \lambda}$.
All of the supersymmetry and the translational symmetry
in one direction are broken by the domain wall solution, however the
R-symmetry of the model is left unbroken.
The quantum fluctuations about the wall solution include the zero mode
Nambu-Goldstone and Goldstino excitations. The
spectrum in addition contains a number of localized massive excitations
corresponding to breathing modes of the wall with masses between
$\sqrt{\mu^2 \lambda}$ and $2\sqrt{\mu^2 \lambda}$, and a continuum of
non-localized modes starting at $2 \sqrt{\mu^2 \lambda}$.
The parameters of the effective domain wall
world volume theory valid below the scale  $\sqrt{\mu^2 \lambda}$ of all the
massive modes are in principle determined by integrating out these massive
excitations.  However, the form of the low energy effective
action is determined solely by the group theoretical
non-linearly realized broken symmetry techniques discussed above.  The thin
domain wall action is given by equations (\ref{ngav-action}), (\ref{AV}) and
(\ref{NG}). Non-BPS domain walls also have been considered in case
one of the target space dimensions is compact \cite{Hou:1999rv}.

The brane localized matter fields'  action is
constructed using the covariant derivatives of the nonlinearly realized
spontaneously broken D=4 super-Poincar\'e symmetries. To this end the
method to include matter fields in theories with non-linearly
realized supersymmetry \cite{Clark:1996aw} is extended to also
include nonlinearly realized translation symmetry. The form of the
leading terms in the domain wall width expansion of the effective action
for scalar, $S(x)$, and fermion, $\psi_i (x)$, matter fields is determined
to be
\be
\Gamma_{\rm matter} = \int d^3 x \det{e}~~{\cal L}_{\rm matter} ,
\ee
with the $G$-invariant matter field lagrangian
\bea
{\cal L}_{\rm matter} &=& {\nabla}_a S \eta^{ab} {\nabla}_b S -V(S) \cr
& &\qquad\quad +i\bar\psi {\gamma}^a {\nabla}_a \psi - m \bar\psi \psi +
Y(S, \bar\psi \psi) .
\eea
The scalar field potential $V(S)$ is an arbitrary function of $S$ and the
generalized Yukawa coupling $Y(S, \bar\psi \psi)$ is a function coupling the
scalar fields $S$ to the scalar bilinears $\bar\psi \psi$.  In the case of a
single species of D=3 Majorana fermion, the Yukawa term terminates at the
form $y(S) \bar\psi \psi$, with the arbitrary Yukawa coupling function
$y(S)$.  The masses and coupling constants of the matter are left as
parameters of the effective theory to be specified by the matching to a
specific underlying domain wall model.  The $G$-covariant derivatives of the
matter fields are obtained in terms of the $G$-covariant space-time
derivatives ${\cal D}_a = e_a^{-1 m}\partial_m$ and the components of the
spin connection $\omega_M^b = \omega^a \Gamma_a^{~b}$ in the $G$-covariant
basis $\omega^a = dx^m e_m^{~a}$
\bea
\nabla_a S &=& {\cal D}_a S \cr
\nabla_a \psi_i &=& {\cal D}_a \psi_i -\frac{i}{2}\Gamma_a^{~b}
\gamma_{bij}\psi_j .
\eea

The fully covariant derivatives ${\cal D}_a$ can be expressed in terms of
the partially covariant Akulov-Volkov derivatives ${\hat{\cal D}}_a$ with
the help of the Nambu-Goto dreibein
\be
{\cal D}_a = N_a^{-1 b} {\hat{\cal D}}_b .
\ee
Likewise, the components of the spin connection can be expressed in this
partially covariant basis $\Gamma_a^{~b} = N_a^{-1 c}
\hat{\Gamma}_c^{~b}$.  So doing, the fully $G$-invariant matter field
lagrangian in the partially covariant basis becomes
\bea
{\cal L}_{\rm matter} &=& \hat{\nabla}_a S n^{ab} \hat{\nabla}_b S -V(S) \cr
& &\qquad\quad +i\bar\psi {\gamma}^b N_b^{-1 a} \hat{\nabla}_a \psi - m
\bar\psi \psi + Y(S, \bar\psi \psi) ,
\eea
with the Nambu-Goto metric given in terms of the Nambu-Goto dreibein $n^{ab}
= N_c^{-1 a} \eta^{cd} N_d^{-1 b}$ while the partially covariant matter
field derivatives are defined by
\bea
\hat{\nabla}_a S &=& \hat{\cal D}_a S \cr
\hat{\nabla}_a \psi_i &=& \hat{\cal D}_a \psi_i
-\frac{i}{2}\hat{\Gamma}_a^{~n} \gamma_{nij}\psi_j .
\eea

Finally, Appendix A is a summary of D=3 Lorentz spinor and tensor
definitions and identities along with Dirac matrix conventions.

\pagebreak

\newsection{\large Coset Construction And Super-Poincar\'{e} Symmetries}

Besides the space-time translation and Lorentz rotation generators, the N=1,
D=4 super-Poincar\'e transformations include the Weyl spinor supersymmetry
charges $Q_\alpha$ and $\bar{Q}_{\dot\alpha}$ obeying the anticommutation
relation
\be
\{ Q_\alpha , {\bar{Q}}_{\dot\alpha}\}= +2\sigma^\mu_{\alpha\dot\alpha}P_\mu
.
\ee
This relation is invariant under the automorphism generated by the $R$
charge
\bea
[R, Q_\alpha ] &=& +Q_\alpha \cr
[R, \bar{Q}_{\dot\alpha} ] &=& -\bar{Q}_{\dot\alpha} \cr
[R , P^\mu ] &=& 0 = [R, M^{\mu\nu} ] .
\eea
A domain wall spontaneously breaks the D=4 symmetries to those of D=3.
In the case of a static planar non-BPS domain wall centered on the $x$-$y$
plane, the above symmetries are broken to only retain those of the D=3
Poincar\'{e} transformations of the wall.  Nambu-Goldstone zero mode degrees
of freedom corresponding to the broken $z$ direction translation generator
and the four supersymmetry generators propagate along the wall.
Geometrically this describes the embedding of a non-BPS spatial 2-brane
having a D=3 space-time world volume into a target N=1, D=4 superspace.  In
the static gauge, the Nambu-Goldstone boson describes motion of the brane in
the spatial ($z$) direction normal to the brane while the Goldstino fields
correspond to motion of the brane in the Grassmann coordinate directions of
N=1, D=4 superspace.

Since the unbroken symmetries are those of the D=3 Poincar\'e group, it is
useful to express the D=4 charges in terms of their D=3 Lorentz group
transformation properties.  However, the SUSY is completely broken in the
non-BPS case, so the fields will not belong to linear SUSY representation
multiplets.  Thus, the space-time translation generator $P^\mu$, which
transforms as a vector
$(\frac{1}{2}, \frac{1}{2})$ representation of the D=4 Lorentz group,
consists of a D=3 Lorentz group vector, $p^m = P^m$, with $m=0,1,2$, and
a D=3 central charge scalar, $Z\equiv P_3$.  Likewise, the
Lorentz transformation charges $M^{\mu\nu}$ are in
the D=4 $(1,1)_A$ representation which consists of two D=3 vector
representations: $M^{mn}=\epsilon^{mnr} M_r$ and $K^m \equiv M^{m3}$.
The $R$ charge is a singlet from both points of view.  Finally the D=4
SUSY $(\frac{1}{2}, 0)$ spinor $Q_\alpha$ and the $(0,\frac{1}{2})$ spinor
$\bar{Q}_{\dot\alpha}$ consist of two D=3 two-component Majorana spinors:
$q_i$
and $s_i$, with $i=1,2$, comprising the charges for (centrally extended)
N=2, D=3 SUSY.  These
spinors are given as linear combinations of $Q_\alpha$ and
$\bar{Q}_{\dot\alpha}$ according to
\bea
q_1 &=& \frac{1}{2}\left[ a( {Q_{1}} - {Q_{2}} ) + \bar{a} (\bar{Q}_1 -
\bar{Q}_2 ) \right]  \cr
& & \cr
q_2 &=& \frac{-i}{2}\left[ a(Q_1 + Q_2 ) - \bar{a} (\bar{Q}_1 + \bar{Q}_2 )
\right]  \cr
& & \cr
s_1 &=& \frac{1}{2}\left[ \bar{a}( {Q_{1}} - {Q_{2}} ) + a (\bar{Q}_1 -
\bar{Q}_2 ) \right]  \cr
& & \cr
s_2 &=& \frac{-i}{2}\left[ \bar{a}(Q_1 + Q_2 ) - a (\bar{Q}_1 + \bar{Q}_2 )
\right]  ,
\eea
where the complex number $a$ is $a\equiv e^{i\pi /4}$.

The N=1, D=4 super-Poincar\'{e} algebra can be written in terms of the
D=3 Lorentz group representation charges as that in equation
(\ref{Poincarealg0}) and the commutators involving the supersymmetry charges
\begin{center}
\begin{tabular}{ll}
$[M^{mn} , q_i ] = -\frac{1}{2} \gamma^{mn}_{ij} q_j\;\;\;\;\;\;\;\;\;$ &
$[K^m , q_i ] = +\frac{1}{2} \gamma^m_{ij} s_j$\\
$[M^{mn} , s_i ] = -\frac{1}{2} \gamma^{mn}_{ij} s_j$  &
$[K^m , s_i ] = -\frac{1}{2} \gamma^m_{ij} q_j$\\
& \\
$[R , q_i ] = +i s_i$ &
$\{ q_i , q_j \} = +2\left(\gamma^m C \right)_{ij} p_m$ \\
$[R , s_i ] = -i q_i$ &
$\{ s_i , s_j \} = +2\left(\gamma^m C \right)_{ij} p_m$\\
& $\{ q_i , s_j \} = -2i C_{ij} Z.$ \\
\end{tabular}
\end{center}
\be
~
\ee
The charge conjugation matrix and the $2+1$ (D=3) dimensional gamma matrices
in the appropriate representation are presented in Appendix A.

The action for the 2-brane can be found by means of the
coset construction.  Towards this end a coset $G/SO(1, 2) \otimes R$ element
$\Omega$, with $G$ the N=1, D=4 super-Poincar\'e group, is written as
\be
\Omega \equiv e^{ix^m p_m } e^{i\left[ \phi Z +\bar\theta_i q_i
+\bar\lambda_i s_i \right]} e^{iv^m K_m} ,
\ee
where the $x^m$ denote the D=3 space-time coordinates parameterizing the
world volume of the 2-brane in the static gauge, while the Nambu-Goldstone
fields, denoted by $\phi (x) , \theta_i (x), \lambda_i (x)$ and $v^m (x)$,
describe the target space excitations of the brane.  Taken together, they
act as coordinates of the coset manifold.  The unbroken symmetry group $H$
is generated by the set of charges $\{p^m, M^m, R \}$.  Multiplication of
the coset elements $\Omega$ by group elements $g \in G$ from the left
results in transformations of the space-time coordinates and the
Nambu-Goldstone fields according to the general structure
\be
g \Omega = \Omega^\prime h ,
\ee
where the infinitesimal transformations $g$ are parameterized as
\be
g = e^{i[ a^m p_m + \bar\xi q + \bar\eta s + zZ + b^m K_m + \alpha^m M_m +
\rho R]} .
\ee
Upon application of the Baker-Campbell-Hausdorff formulae for infinitesimal
$A$
and arbitrary $B$, with Lie derivative ${\cal L}_A \cdot B = [A,B]$,
\bea
e^A e^B e^{-A} &=& e^{B + [A,B]}  \cr
e^B e^A &=& e^{B + {\cal L}_{B/2}\cdot \left[ A + \coth{({\cal L}_{B/2})}
\cdot
A \right]} \cr
e^A e^B &=& e^{B - {\cal L}_{B/2}\cdot \left[ A - \coth{({\cal L}_{B/2})}
\cdot
A \right]} ,
\eea
the transformed coset element is given by the total variation of the fields
so that
\be
\Omega^\prime = e^{i x^{\prime m} p_m}
e^{i \left[ \phi^\prime (x^\prime) Z +  \bar\theta_i^\prime (x^\prime)
q_i + \bar\lambda^\prime_i (x^\prime) s_i \right]}
e^{i  v^{\prime n} (x^\prime) K_n } ,
\ee
while $h$ allows $\Omega^\prime$ to be written as a coset element and is
given by
\be
h = e^{i\left(\alpha^n M_m + \rho R - \frac{1}{2}
\frac{\tanh{\frac{1}{2}\sqrt{v^2}}}{\frac{1}{2}\sqrt{v^2}} b_m v_r
\epsilon^{mrn} M_n \right)} =e^{i\rho R} e^{i\beta^m (g, v) M_m} .
\label{h}
\ee
The infinitesimal transformations induced on the 2-brane space-time
coordinates and fields are obtained as
\bea
x^{\prime m} &=& x^m + a^m  - i (\bar\xi \gamma^m \theta + \bar\eta \gamma^m
\lambda ) -\phi b^m + \epsilon^{mnr} \alpha_n x_r \cr
\Delta\theta_i &=& \xi_i + \frac{i}{2} b_m (\gamma^m \lambda )_i - i\rho
\lambda_i  -\frac{i}{2} \alpha_m (\gamma^m \theta )_i \cr
\Delta\lambda_i &=& \eta_i - \frac{i}{2} b_m (\gamma^m \theta )_i + i\rho
\theta_i  -\frac{i}{2} \alpha_m (\gamma^m \lambda )_i \cr
\Delta \phi &=& z + (\xi \gamma^0 \lambda - \theta \gamma^0 \eta ) - b^m x_m
\cr
\Delta v^m &=& +\frac{\sqrt{v^2}}{\tanh{\sqrt{v^2}}} \left( b^m -
\frac{v^r b_r v^m}{v^2} \right) + \frac{v^r b_r v^m}{v^2} + \epsilon^{mnr}
\alpha_n v_r .
\eea
The intrinsic variation of the fields, $\delta \varphi \equiv
\varphi^\prime (x) -\varphi (x)$, is related to the above total variation,
$\Delta \varphi $, by the Taylor expansion shift in the space-time
coordinates:
\be
\delta \varphi = \Delta \varphi - \delta x^m \partial_m \varphi ,
\ee
with $\delta x^m = x^{\prime m} - x^m $.

The nonlinearly realized D=4 super-Poincar\'e symmetries induce a field
dependent general coordinate transformation of the world volume space-time
coordinates.  From above, the general coordinate transformation for the
world volume space-time coordinate differentials is given by
\be
dx^{\prime m} = dx^n G_n^{~m} ,
\ee
where
\bea
G_n^{~m} &=& \frac{\partial x^{\prime m}
}{\partial x^n} \cr
&=& \delta_n^{~m} - i (\partial_n \theta \gamma^0 \gamma^m \xi + \partial_n
\lambda \gamma^0 \gamma^m \eta ) -\partial_n \phi b^m
+ \epsilon_n^{~~ms} \alpha_s .
\eea
The $G$-invariant interval can be formed by means of the metric tensor
$g_{mn}$ so that $ds^2 = dx^m g_{mn} dx^n = ds^{\prime 2} = dx^{\prime m}
g^\prime_{mn} dx^{\prime n}$ where the metric transforms as a tensor
\be
g^\prime_{mn} = G^{-1 r}_m g_{rs} G^{-1 s}_n .
\label{gprime}
\ee
The metric can be constructed from the domain wall dreibein obtained from
the Maurer-Cartan one-form.

\pagebreak

\newsection{\large Maurer-Cartan One-Forms And The Invariant Action}

According to the coset construction method, the dreibein, the covariant
derivatives of the Nambu-Goldstone fields and the spin connection can be
obtained from the Maurer-Cartan one-forms.  The Maurer-Cartan one-forms can
be determined by use of the Feynman formula for the variation of an
exponential operator along with the B-C-H formula $e^A B e^{-A} = e^{{\cal
L}_A} \cdot B$,
\bea
e^{-iA} \delta e^{+iA} &=& \int_0^1 dt e^{-itA} (i\delta A) e^{+itA} \cr
&=& \frac{e^{{\cal L}_{-iA}} -1}{{\cal L}_{-iA}} \cdot (i\delta A)  \cr
&=& i\delta A - \frac{(-i)^2}{2} [A, \delta A] - \cdots \cr
& & \qquad - \frac{i^{(n+1)}}{(n+1)!} \underbrace{[A,[A,\ldots [A, \delta A]
\cdots ]}_{\rm n-commutators} - \cdots .
\eea
The Maurer-Cartan one-forms are given as
\be
\Omega^{-1} d\Omega = i\left[ \omega^{a} p_a +\bar\omega_{qi} q_i
+\bar\omega_{si}
s_i + \omega_Z Z+ \omega_{K}^a K_a + \omega_{M}^a M_a + \omega_R R \right]
\ee
where the individual world volume one-forms are found to be
\bea
\omega^{a} &=& \left( dx^m +i d\theta \gamma^0 \gamma^m \theta +i d\lambda
\gamma^0 \gamma^m \lambda \right) \left( \delta_{m}^{~a} +(\cosh{\sqrt{v^2}}
-1)\frac{v_m v^a}{v^2} \right)  \cr
& & \cr
& & +\left( d\phi +  d\theta \gamma^0 \lambda - d\lambda \gamma^0 \theta
\right) \frac{\sinh{\sqrt{v^2}}}{\sqrt{v^2}}v^a  \cr
\omega_{qi} &=& \cosh{\frac{1}{2}\sqrt{v^2}} d\theta_i
-\frac{i}{2}\frac{\sinh{\frac{1}{2}\sqrt{v^2}}}{\frac{1}{2}\sqrt{v^2}}
(\rlap{/}{v} d\lambda)_i  \cr
\omega_{si} &=& \cosh{\frac{1}{2}\sqrt{v^2}} d\lambda_i +
\frac{i}{2}\frac{\sinh{\frac{1}{2}\sqrt{v^2}}}{\frac{1}{2}\sqrt{v^2}}
(\rlap{/}{v} d\theta )_i  \cr
\omega_Z &=& \left(d\phi +  d\theta \gamma^0 \lambda - d\lambda \gamma^0
\theta \right) \cosh{\sqrt{v^2}} \cr
& & + \left( dx^m +i d\theta \gamma^0 \gamma^m \theta +i d\lambda
\gamma^0 \gamma^m \lambda \right) v_m \frac{\sinh{\sqrt{v^2}}}{\sqrt{v^2}}
\cr
\omega_{K}^a &=& dv^b \left[ \delta_b^{~a} + \left(
\frac{\sinh{\sqrt{v^2}}}{\sqrt{v^2}}
-1 \right) P_{vTb}^{~~~a} \right] \cr
\omega_{M}^a &=& \left( \cosh{\sqrt{v^2}} - 1 \right) \frac{v_b dv_c}{v^2}
\epsilon^{abc} \cr
\omega_R &=& 0 .
\label{oneforms}
\eea

The Maurer-Cartan one-forms transform covariantly under all of the $G$
symmetries
except the unbroken D=3 Lorentz transformation one-form $\omega_{M}^a$,
which
transforms with an additional shift ($hdh^{-1} \neq 0$) under the broken D=4
Lorentz transformations, as required of a connection one-form.  Explicitly,
recalling that left multiplication by a
group member induces a transformation in the world volume space-time
coordinates and fields, $g\Omega =
\Omega^\prime h$, the Maurer-Cartan one-forms transform as
\be
\left( \Omega^{-1}d \Omega \right)^\prime = h (\Omega^{-1} d\Omega )h^{-1}
+hdh^{-1} .
\ee
From this the dreibein, the covariant derivatives and the spin connection
transformations can be obtained.  In addition, as shown below, the
$G$-covariant one-form $\omega_Z$ can be used to eliminate the would be
Nambu-Goldstone field $v^m$ so that the independent degrees of freedom
include only the Nambu-Goldstone modes for the spontaneously broken
translation symmetry and supersymmetry.  These correspond to excitations of
the 2-brane into N=1, D=4 superspace directions \lq\lq normal"  to the
spatial non-BPS domain wall brane.  Towards this end, the world volume
tangent space covariant coordinate basis differentials are given by the
$\omega^{a}$ one-form.  For a $G$-transformation they transform under the
broken D=4 Lorentz transformations and the unbroken D=3 Lorentz rotations
according to their D=3 (local) Lorentz vector nature as given by $h$ in
equation (\ref{h}) (and are $R$ invariant).  Writing $h$ as $h= e^{i\beta_a
(g, v) M^a}$, the transformation of $\omega^a$ is given by
\be
\omega^{\prime a} = \omega^b L_b^{~a} ,
\label{structuretrans}
\ee
where the transformation is simply
\be
L_b^{~a} = \delta_b^{~a} + \beta^c \epsilon_{cb}^{~~a} = \left(e^{-i\beta_c
\tilde{M}_{\rm vector}^c}\right)_b^{~a} ,
\label{L}
\ee
with the D=3 Lorentz vector representation matrix $(\tilde{M}_{{\rm
vector}~c})_a^{~~b} = i
\epsilon_{ca}^{~~~b}$.  The determinant of $L$ is unity: $\det{L} = 1$.
(The remaining one-forms similarly transform according to their D=3 Lorentz
character.  Because of this local Lorentz structure group transformation
property of the vector one-forms, their indices are denoted by letters from
the beginning of the alphabet: $a, b, c, \ldots ~=0, 1, 2$.)

The two sets of coordinate basis differentials $dx^m$ and $\omega^a$ are
related to each other through the dreibein $e_m^{~a}$
\be
\omega^a = dx^m e_m^{~a} .
\ee
From equation (\ref{oneforms}) this yields
\bea
e_m^{~a} &=& \left(\delta_m^{~b} + i\partial_m \theta \gamma^0 \gamma^b
\theta +\partial_m \lambda \gamma^0 \gamma^b \lambda \right) \left(
\delta_b^{~a} + \left( \cosh{\sqrt{v^2}} -1 \right) \frac{v_b
v^a}{v^2}\right. \cr
& &\left. \qquad +\left( \hat{\cal D}_b \phi + \hat{\cal D}_b \theta
\gamma^0 \lambda - \theta \gamma^0 \hat{\cal D}_b \lambda \right) v^a
\frac{\sinh{\sqrt{v^2}}}{\sqrt{v^2}} \right) ,
\label{e2}
\eea
with the Akulov-Volkov derivative $\hat{\cal D}_a = \hat{e}_a^{-1
m}\partial_m$ defined below (see equation (\ref{avderiv2})).  Under a
$G$-transformation the dreibein transforms with one world index and one
tangent space (structure group) index as
\be
e_m^{\prime ~a}= G_m^{-1 n} e_n^{~b} L_b^{~a} ,
\ee
and likewise for the inverse dreibein $e^{\prime -1 m}_a = L_a^{-1 b} 
e_b^{-1 n} G_n^{~ m}$.
By direct calculation from the form of $L$, equation (\ref{L}), the flat
tangent space metric, $\eta_{ab}$, is invariant
\be
\eta_{ab}^\prime = L_a^{~c} \eta_{cd} L_b^{~d} = \eta_{ab} .
\ee
The metric tensor is given in terms of the dreibein as
\be
g_{mn} = e_m^{~a} \eta_{ab} e_n^{~b} ,
\ee
the transformation properties of which are given by equation (\ref{gprime})
and follow from those of the dreibein and the flat tangent space metric.
Consequently the covariant Maurer-Cartan one-form can be used to express the
invariant interval as
\be
ds^2 = dx^m g_{mn} dx^n = \omega^a \eta_{ab} \omega^b .
\ee
The leading term in the D=4 super-Poincar\'e invariant action is given by
the \lq\lq cosmological constant\rq\rq  term
\be
\Gamma = -\sigma \int d^3 x \det{e} ,
\ee
with the brane tension parameter $\sigma$.  The lagrangian is the constant
brane tension integrated over the area of the brane.  The action is
invariant
\be
\Gamma^\prime = -\sigma \int d^3 x^\prime \det{e^\prime} = -\sigma \int
\left( d^3 x \det{G} \right) \left( \det{G^{-1}}\det{e} \det{L} \right) =
-\sigma \int d^3 x \det{e} = \Gamma .
\ee

\pagebreak

\newsection{\large Dreibeine, Covariant Derivatives And Brane Dynamics}

The world volume exterior derivative, $d = dx^m \partial_m$,
can also be written in terms of the fully $G$-covariant one-form basis
\be
d = dx^m \partial_m  = \omega^a e_a^{-1 m} \partial_m \equiv \omega^a
{\cal D}_a ,
\ee
with the fully $G$-covariant derivative
\be
{\cal D}_a \equiv e_a^{-1 m} \partial_m .
\ee
The exterior derivative is fully $G$-invariant $d^\prime = d$ since the
coordinate derivative transforms inversely to the coordinate differential: $
\partial_m^\prime = G_m^{-1 n} \partial_n $.  Hence, ${\cal D}_a$ transforms
as ${\cal D}^\prime_a = L_a^{-1 b} {\cal D}_b $.

For each one-form, $\omega_{Q_{\varphi}}$, with $Q_{\varphi} = \{ q_i, s_i,
Z, K^m \}$, the respective covariant derivative of the related
Nambu-Goldstone field, $\varphi = \{\theta_i , \lambda_i , \phi, v^m \}$, is
defined according to
\bea
\omega_{Q_{\varphi}} &\equiv & \omega^a {\nabla}_a \varphi =dx^m e_m^{~a}
{\nabla}_a \varphi \cr
&\equiv & dx^m \omega_{Q_{\varphi} m} = \omega^a e_a^{-1 m}
\omega_{Q_{\varphi}
m} .
\eea
Hence it is obtained that
\be
{\nabla}_a\varphi = e_a^{-1 m} \omega_{Q_{\varphi} m}
\ee
or the inverse
\be
e_m^{~a} {\nabla}_a \varphi = \omega_{Q_{\varphi} m} .
\ee
Recall that each one-form begins with the space-time derivative of the
associated
Nambu-Goldstone field: $\omega_{Q_{\varphi} m} =\partial_m \varphi +
\cdots$.

Besides the fully $G$-covariant basis of one-forms, partially covariant
bases associated with restricted motions in the coset manifold can be
defined.  In particular the basis obtained from motion in the manifold with
the coset coordinate $v^n =0$ is a D=3 Lorentz but not D=4 Lorentz covariant
one-form basis.  Most directly these one-forms, dreibein and partially
covariant derivatives can be obtained by taking the $v^n$ field to zero in
the above expressions, for example, $\hat\omega^a \equiv \omega^a \vert_{v^n
= 0}$.  Alternatively, since the Maurer-Cartan one-forms can be built-up
sequentially by including the different symmetry generators
\be
\Omega^{-1} d\Omega = e^{-iv^n K_n} [ d+ \hat\Omega^{-1} d\hat\Omega ]
e^{+iv^n K_n} ,
\ee
where the $\hat\Omega$ includes the remaining generators, the partially
covariant one-forms are given by
\bea
\hat\Omega^{-1} d \hat\Omega &\equiv & i\left[ \hat\omega^{a} p_a
+\bar{\hat\omega}_{qi} q_i +\bar{\hat\omega}_{si}
s_i + \hat\omega_Z Z \right] \cr
&=& i[dx^a +id\theta \gamma^0 \gamma^a \theta +i d\lambda \gamma^0 \gamma^a
\lambda ]p_a +id\bar\theta q +id\bar\lambda s \cr
& & +i[ d\phi + d\theta \gamma^0 \lambda - d\lambda \gamma^0 \theta ]Z .
\label{hatforms}
\eea

The space-time coordinate differentials can be expressed in terms of this
one-form basis through the Akulov-Volkov dreibein
\be
\hat\omega^a = dx^m \hat{e}_m^{~a}.
\label{partial1form}
\ee
From equations (\ref{hatforms}) and (\ref{partial1form}), (or
$\hat{e}_m^{~a} = e_m^{~a} \vert_{v^n = 0}$ in equation (\ref{e2}))
\be
\hat{e}_{m}^{~~a} = A_m^{~~a}
\ee
where the Akulov-Volkov matrix $A_m^{~a}$ is
defined as
\be
A_m^{~a} = \delta_m^{~a} + i \partial_m \theta \gamma^0 \gamma^a \theta
+ i \partial_m \lambda \gamma^0 \gamma^a \lambda .
\ee
From these the D=3 SUSY SO(1,2) covariant derivatives follow
\be
d = dx^m \partial_m \equiv \hat\omega^a \hat{\cal D}_a ,
\ee
whence $\hat{\cal D}_a = \hat{e}_{a}^{-1 ~m} \partial_m$
where the inverse dreibein $\hat{e}_{a}^{-1~m}$, so that $\hat{e}_{a}^{-1~m}
\hat{e}_{m}^{~~b}= \delta_a^{~b}$ and $\hat{e}_{m}^{~~a} \hat{e}_{a}^{-1~n}
= \delta_m^{~n}$, is given by
\be
\hat{e}_{a}^{-1~~m} = A_a^{-1~m} .
\ee
Hence the partial covariant Akulov-Volkov derivative is obtained
\be
\hat{\cal D}_a = \hat{e}_{a}^{-1~m}\partial_m = A_a^{-1~m} \partial_m .
\label{avderiv2}
\ee

The $G$-transformation properties of the partially covariant one-forms,
equation (\ref{hatforms}), can be found from the factorization of the coset
element and the general transformation law.  Writing $\Omega = \hat{\Omega}
\Omega_K$, the transformation law $g\Omega = \Omega^\prime h$
implies that
\be
\left( \hat{\Omega}^{-1} d \hat{\Omega}\right)^\prime = \hat{h} \left(
\hat{\Omega}^{-1} d \hat{\Omega}\right) \hat{h}^{-1} ,
\ee
where now $\hat{h}$ involves the broken and unbroken Lorentz generators but
with the field independent transformation parameters of $g$, $\hat{h} =
e^{i\rho R} e^{ib^n K_n} e^{i\alpha^n M_n} $.  In particular, this yields
the non-covariant transformation law for $\hat{\omega}^a$ (even so, the use
of indices from the beginning of the alphabet is retained)
\bea
\hat{\omega}^{\prime a} &=& \hat{\omega}^b \hat{L}_b^{~a} \cr
&=& \hat{\omega}^b \left( \delta_b^{~a} + \alpha^c \epsilon_{cb}^{~~~a} -
\hat\nabla_b \phi b^a \right) ,
\eea
with the partially covariant derivative of $\phi$, $\hat\nabla_a \phi$,
given by the Maurer-Cartan one-form $\hat\omega_Z$,
\be
\hat\omega_Z = \hat\omega^a \hat\nabla_a \phi = \hat\omega^a \left[\hat{\cal 
D}_a
\phi + \hat{\cal D}_a \theta \gamma^0 \lambda -\theta \gamma^0 \hat{\cal 
D}_a \lambda
\right] .
\ee
Hence, the Akulov-Volkov derivative transforms as
\be
\hat{\cal D}_a^\prime = \hat{L}_a^{-1 b} \hat{\cal D}_b ,
\ee
and as such is not fully $G$-covariant due to its variation under the broken
Lorentz transformations ($b^n \neq 0$).  As with $\hat{\omega}^a$, the
Akulov-Volkov derivative $\hat{\cal D}_a$ is only $SO(1, 2)$ partially
covariant.

It is useful to expand the one-forms of the fully $SO(1, 3)$ covariant basis
in terms of the $SO(1, 2)$ covariant basis.  The two bases are related as
\bea
\hat\omega^a &=& \omega^b e_b^{-1~m} \hat{e}_m^{~~a} \cr
\omega^a &=& \hat\omega^b \hat{e}_b^{-1~m} e_m^{~a}
\label{changebasis}
\eea
which follows from the superspace coordinate differentials
\be
dx^m = \omega^a e_a^{-1~m} = \hat\omega^a \hat{e}_a^{-1~m} .
\ee
Likewise, through the exterior derivative, $d=dx^m \partial_m = \omega^a
{\cal D}_a = \hat\omega^a \hat{\cal D}_a$, the (partially) covariant
derivatives are related
\bea
\hat{\cal D}_a &=& \hat{e}_a ^{-1~m}e_m^{~b} {\cal D}_b \cr
{\cal D}_a &=& e_a^{-1~m}\hat{e}_m^{~b} \hat{\cal D}_b .
\label{relatederiv}
\eea

In particular the $G$-covariant coordinate differential one-form,
$\omega^a$, has a simple relation to the partially covariant coordinate
differential one-form, $\hat\omega^a$,
\be
\omega^a = \hat{\omega}^b N_b^{~a} ,
\ee
where the Nambu-Goto dreibein, $N_b^{~a}$, is found from equations
(\ref{oneforms}) and (\ref{hatforms})
\bea
N_b^{~a} &=& \delta_b^{~a} + \left( \cosh{\sqrt{v^2}} -1 \right) \frac{v_b
v^a}{v^2} \cr
& & + \left( \hat{\cal D}_b \phi + \hat{\cal D}_b \theta \gamma^0 \lambda
-\theta \gamma^0 \hat{\cal D}_b \lambda \right) v^a
\frac{\sinh{\sqrt{v^2}}}{\sqrt{v^2}} .
\label{ngdreibein2}
\eea
Similarly the dreibeine are related, equation (\ref{e2}),
\be
e_m^{~a} = \hat{e}_m^{~b} N_b^{~a} .
\ee
Thus the invariant action takes on a factorized form
\be
\Gamma = -\sigma \int d^3 x \det{e} = -\sigma \int d^3 x \det{\hat{e}}
\det{N} .
\ee
The $\det{\hat{e}}$ has the usual form of the Akulov-Volkov determinant for
spontaneously broken N=2, D=3 supersymmetry.  The $\det{N}$ term can be
evaluated to yield the SUSY generalization to the Nambu-Goto action for the
p=2 brane allowing for its motion into the Grassmann directions of the
target N=1, D=4 superspace
\be
\det{N} = \cosh{\sqrt{v^2}} \left[ 1 + \left(\hat{\cal D}_a \phi + \hat{\cal
D}_a \theta \gamma^0 \lambda -\theta \gamma^0 \hat{\cal D}_a \lambda \right)
v^a \frac{\tanh{\sqrt{v^2}}}{\sqrt{v^2}} \right].
\label{detfullN}
\ee

There are two equivalent ways in which to proceed in order to simplify the
action by the elimination of the $v^m$ field.  The Euler-Lagrange approach
is a result of the fact that the action depends only on $v^m$ and not its
derivatives.  Hence the $v^m$ field equation, $\delta \Gamma / \delta v^m =
0$, will express $v^m$ in terms of the independent Nambu-Goldstone fields,
$\phi$, $\theta$ and $\lambda$.  Alternatively, the Maurer-Cartan one-form
associated with the broken translation generator $Z$ can be $G$-covariantly
set to zero.  Expanding the $\omega_Z$ one-form in terms of the
$\hat\omega^a$ basis gives
\be
\omega_Z = \hat\omega^a \cosh{\sqrt{v^2}} \left[ \left( \hat{\cal D}_a \phi
+ \hat{\cal D}_a \theta \gamma^0 \lambda -\theta \gamma^0 \hat{\cal D}_a
\lambda \right)   + v_a \frac{\tanh{\sqrt{v^2}}}{\sqrt{v^2}}\right] .
\ee
Setting this to zero results in the \lq\lq inverse Higgs mechanism\rq\rq
\be
v_a \frac{\tanh{\sqrt{v^2}}}{\sqrt{v^2}} = -\left( \hat{\cal D}_a \phi +
\hat{\cal D}_a \theta \gamma^0 \lambda -\theta \gamma^0 \hat{\cal D}_a
\lambda \right)= - \hat\nabla_a \phi  .
\ee
This result is also obtained in the Euler-Lagrange approach.
Substituting this into the determinant of the Nambu-Goto dreibein yields the
SUSY generalization of the Nambu-Goto lagrangian as given in equation
(\ref{NG})
\bea
\det{N} &=& \frac{1}{\cosh{\sqrt{v^2}}} \cr
&=& \sqrt{1- \left({\hat{\cal D}}_a \phi +   {\hat{\cal D}}_a \theta
\gamma^0 \lambda - \theta \gamma^0 {\hat{\cal D}}_a \lambda \right)^2} .
\eea
Hence the complete $G$-invariant Nambu-Goto-Akulov-Volkov action is given by
\bea
\Gamma &=& -\sigma \int d^3 x \left\{ \det{\left[ \delta_m^{~a} + i
\partial_m \theta \gamma^0 \gamma^a \theta
+ i \partial_m \lambda \gamma^0 \gamma^a \lambda \right]} \times \right. \cr
& &\left. \qquad\qquad \times \sqrt{1- \left({\hat{\cal D}}_b \phi +
{\hat{\cal D}}_b \theta \gamma^0 \lambda - \theta \gamma^0 {\hat{\cal D}}_b
\lambda \right)^2}\right\} .
\eea

Returning to equation (\ref{detfullN}) and treating all fields as
independent leads to the $\phi$ equation of motion as the D=3 Bianchi
identity for the field strength vector $F^m$
\be
0=\frac{\delta \Gamma}{\delta \phi} = \partial_m F^m ,
\ee
where
\be
F^m = \det{\hat{e}} v^a \hat{e}^{-1 m}_a
\frac{\sinh{\sqrt{v^2}}}{\sqrt{v^2}}  .
\ee
Substituting this back into the lagrangian yields
\be
\det{\hat{e}}\det{N} = \det{\hat{e}} \cosh{\sqrt{v^2}} + F^m \left[
\partial_m \phi + \partial_m \theta \gamma^0 \lambda - \theta \gamma^0
\partial_m \lambda \right] .
\ee
Exploiting the definition of $F^m$ so that
\be
\frac{v^a v^b}{v^2} = \frac{(F^m \hat{e}_m^{~a})(F^n
\hat{e}_n^{~b})}{(F\hat{e})^2}
\ee
results in
\be
\cosh{\sqrt{v^2}} = \sqrt{\left( 1+
\frac{(F\hat{e})^2}{(\det{\hat{e}})^2}\right)}.
\ee
Integrating this over the world volume, the non-BPS p=2 brane
supersymmetric Nambu-Goto-Akulov-Volkov action is dual to the D2-brane
supersymmetric Born-Infeld action
\bea
\Gamma &=& -\sigma \int d^3 x \left(\sqrt{ \det{\left( \hat{g}_{mn} +
F_{mn}\right)}} +F^n [\partial_n \theta \gamma^0 \lambda -\theta \gamma^0
\partial_n \lambda ]\right) \cr
&=& -\sigma \int d^3 x \left(\sqrt{ \det{\hat{g}} + F^m \hat{g}_{mn} F^n }
+F^n [\partial_n \theta \gamma^0 \lambda -\theta \gamma^0 \partial_n \lambda
]\right),
\eea
where the Akulov-Volkov metric is given by $\hat{g}_{mn} =
\hat{e}_m^{~a}\eta_{ab}\hat{e}_n^{~b}$ and $F_{mn} = \epsilon_{mnr} F^r$.
\newpage

\newsection{\large Brane Localized Matter Fields}

The matter fields localized on the brane are characterized by their D=3
Lorentz transformation properties.  A scalar field, $S(x)$, is in the
trivial representation of the Lorentz group: $M^a \rightarrow
(\tilde{M}^a)=0$.  Fermion fields, $\psi_i (x)$, are in the spinor
representation: $M^a \rightarrow (\tilde{M}^a)_{ij}=-1/2 \gamma^a_{ij}$.  
Each
matter field, $M(x)$, transforms under $G$ as
\be
M^\prime (x^\prime ) \equiv \tilde{h} M(x) ,
\ee
where $\tilde{h}$ is given by $h$, equation (\ref{h}), with $M^a$ replaced 
by
$\tilde{M}^a$ and the field's $R$-weight phase, $\rho_M$, a model dependent
convention
\be
\tilde{h} = e^{i\beta_a (g,v) \tilde{M}^a} e^{i\rho_M} .
\ee
The covariant derivative for the matter field is defined using the spin
connection one-form
\be
\nabla M \equiv ( d +i \omega_{M}^a \tilde{M}_a ) M .
\ee
The transformation properties of the covariant derivative,
\be
(\nabla M)^\prime (x^\prime) = \tilde{h} \nabla M(x) ,
\label{Mprime}
\ee
are obtained from the invariant nature of the exterior derivative $d$, the
field dependent transformation equation for $M$ and the inhomogeneous
transformation property of the connection.  For infinitesimal $G$
transformations recall that
$hdh^{-1} = -i d\beta^a M_a$ so that the connection one-form transforms
according to (with $(\tilde{M}_{\rm vector}^a )_{bc} =i\epsilon^a_{~bc}$)
\bea
\omega^{\prime a}_{M} &=& \omega_M^b L_b^{~a}  - d\beta^{a} \cr
&=& \omega_M^{~b} \left( e^{-i\beta_c \tilde{M}^c_{\rm 
vector}}\right)_b^{~~a}
- d\beta^a .
\eea
The covariant derivative transformation law equation (\ref{Mprime}) follows.

Expanding the covariant derivative one-form in terms of the tangent space
covariant coordinate basis differentials, $\omega^a$, the component form of
the covariant derivative is obtained
\be
\nabla_a M = \left( {\cal D}_a +i \Gamma_a^{~b} \tilde{M}_b \right) M ,
\ee
where $\Gamma_a^{~b}$ are the components of the connection, $\omega_{M}^{~b}
= \omega^a \Gamma_a^{~b}$.  Also, in component form, the connection
transformation law is found to be
\be
\Gamma_a^{\prime~b} = L_a^{-1 c} \Gamma_c^{~d} L_d^{~b} - L_a^{-1 c} {\cal
D}_c \beta^b .
\ee
Since the covariant coordinate differentials transform according to the D=3
(field dependent) local Lorentz (structure) group vector representation
matrices, $L_a^{~b}$, equation (\ref{structuretrans}), the component form of
the covariant derivative has the $G$ transformation law
\be
\left( \nabla_a M\right)^\prime (x^\prime) = \tilde{h} L_a^{-1 b} \nabla_b M
(x) .
\ee

For scalar matter fields the covariant derivative is simply the covariant
space-time derivative ${\cal D}_a = e_a^{-1 m}\partial_m$:
\be
\nabla_a S(x) = {\cal D}_a S(x) .
\ee
Since $S$ is invariant, $S^\prime (x^\prime)= S(x)$, the covariant
derivative transforms as a tangent space vector
\be
(\nabla_a S)^\prime (x^\prime) = L_a^{-1 b} \nabla_b S(x).
\ee
Because the flat tangent space metric, $\eta^{ab}$, is invariant, the
leading terms in the brane width expansion of the $G$-invariant action for
the scalar matter field are obtained as
\be
\Gamma_{S} = \int d^3 x \det{e} ~{\cal L}_S ,
\ee
with the scalar field invariant lagrangian (that is invariant under total
$G$-transformations and hence a scalar density under intrinsic
$G$-transformations) given by
\be
{\cal L}_S = ( \nabla_a S \eta^{ab} \nabla_b S ) -V(S) ,
\ee
where the scalar field potential $V(S)$ is an arbitrary function of $S$.

The fermion matter field $\psi_i (x)$ transforms as the D=3 Lorentz group
spinor representation
\be
\psi^\prime_i (x^\prime) = \tilde{h}_{ij} \psi_j (x) ,
\ee
with (suppressing the $R$-transformation weight)
\be
\tilde{h}_{ij} = \left( e^{-\frac{i}{2} \beta_a \gamma^a }\right)_{ij} .
\ee
Hence the bilinear product, $\bar\psi \psi$, is invariant, $(\bar\psi
\psi)^\prime (x^\prime) = (\bar\psi \psi ) (x)$.
The vector bilinear product transforms as a tangent space vector, $(\bar\psi
\gamma^a \psi)^\prime (x^\prime) = (\bar\psi \gamma^b \psi ) (x) L_b^{~a}$.
The covariant derivative now involves the spin connection and, in component
form, is given by
\be
\nabla_a \psi_i = {\cal D}_a \psi_i -\frac{i}{2} \Gamma_a^{~b} \gamma_{bij}
\psi_j .
\ee
The fermion covariant derivative transforms according to
\be
(\nabla_a \psi_i )^\prime (x^\prime) = \tilde{h}_{ij} L_a^{-1 b} \nabla_b
\psi_j .
\ee
The invariant kinetic energy bilinear is given by $(\bar\psi \gamma^a
\nabla_a \psi)$ so that
\be
(\bar\psi \gamma^a \nabla_a \psi)^\prime (x^\prime) = (\bar\psi \gamma^a
\nabla_a \psi) (x).
\ee
The $G$ invariant action has the form
\be
\Gamma_f = \int d^3 x \det{e}~ {\cal L}_f ,
\ee
where the invariant lagrangian involves the fermion and scalar matter fields
\be
{\cal L}_f = i \bar\psi \gamma^a \nabla_a \psi - m \bar\psi \psi + Y(S,
\bar\psi \psi) ,
\ee
with the generalized Yukawa coupling to the scalar fields, $Y(S, \bar\psi
\psi)$.  In the case of a single species of D=3 Majorana fermion, the Yukawa
term terminates at the form $y(S) \bar\psi \psi$, with the arbitrary Yukawa
coupling function $y(S)$.

The above covariant derivatives were expanded in the fully covariant
$\omega^a$ basis, the relation to the expansion in terms of the partially
covariant $\hat{\omega}^a$ basis can also be obtained.  As found above, the
scalar and fermion covariant derivatives in the fully $G$ covariant basis
are
\bea
\nabla_a S &=& {\cal D}_a S \cr
\nabla_a \psi_i &=& {\cal D}_a \psi_i -\frac{i}{2}\Gamma_a^{~b}
\gamma_{bij}\psi_j .
\eea
The covariant derivatives are related through the exterior derivative and
the dreibeine as in equations (\ref{changebasis})-(\ref{relatederiv}).  The
coordinate differentials are related according to
\bea
\omega^a &=& dx^m e_m^{~a}  \qquad\qquad dx^m = \omega^a e_a^{-1 m} \cr
\hat{\omega}^a &=& dx^m \hat{e}_m^{~a}  \qquad\qquad dx^m = \hat{\omega}^a
\hat{e}_a^{-1 m} \cr
\omega^a &=& \hat{\omega}^b N_b^{~a}  \qquad\qquad  e_a^{-1 n}
\hat{e}_n^{~b} = N_a^{-1 b} .
\eea
The relation between the covariant derivatives is found through the exterior
derivative
\bea
d &=& dx^m \partial_m = \omega^a {\cal D}_a = \hat{\omega}^a \hat{\cal D}_a
\cr
&=& dx^m e_m^{~a} {\cal D}_a = dx^m \hat{e}_m^{~a} \hat{\cal D}_a ,
\eea
with, as previously defined,
\bea
{\cal D}_a &=& e_a^{-1 m} \partial_m \cr
\hat{\cal D}_a &=& \hat{e}_a^{-1 m} \partial_m .
\eea
The relation between (partially) covariant derivatives is secured (see
equation (\ref{relatederiv}))
\bea
{\cal D}_a &=& e_a^{-1 m} \hat{e}_m^{~b} \hat{\cal D}_b \cr
&=& N_a^{-1 b} \hat{\cal D}_b .
\eea
Recall that $\hat{\cal D}_a = A_a^{-1 m} \partial_m$ is just the SUSY
covariant Akulov-Volkov derivative, equation (\ref{avderiv2}).

Besides the derivatives, also the connection can be expressed in terms of
the partially covariant coordinate differentials.  Recall the connection
from equation (\ref{oneforms})
\bea
\omega_M^{~ b} &=& \omega^a \Gamma_a^{~ b} = \hat{\omega}^a
\hat{\Gamma}_a^{~ b} \cr
&=& \left( \cosh{\sqrt{v^2}} -1 \right) \epsilon^{bac} \frac{v_a dv_c}{v^2}
.
\eea
Upon application of the $\omega_Z =0$ constraint, the above expression
becomes
\be
\omega_M^{~ b} = \left( 1/\det{N} -1 \right) \epsilon^{bac} \frac{\varphi_a
d\varphi_c}{\varphi^2} ,
\ee
where $\varphi_a$ is defined by
\be
\varphi_a = \hat{\cal D}_a \phi + \hat{\cal D}_a \theta \gamma^0
\lambda - \theta \gamma^0 \hat{\cal D}_a \lambda . 
\ee
In components in the fully covariant basis, the connection involves the
fully covariant derivative
\be
\Gamma_a^{~b} = \left( \cosh{\sqrt{v^2}} -1 \right) \epsilon^{bcd} \frac{v_c
{\cal D}_a v_d}{v^2} .
\ee
Using the relation between the two bases, the connection components are
related according to
\be
\Gamma_a^{~b} = N_a^{-1 c} \hat{\Gamma}_c^{~b} ,
\ee
with
\be
\hat{\Gamma}_a^{~b} = \left( \cosh{\sqrt{v^2}} -1 \right) \epsilon^{bcd}
\frac{v_c \hat{\cal D}_a v_d}{v^2} ,
\ee
which, after imposition of the $\omega_Z = 0$ constraint, becomes
\be
\hat{\Gamma}_a^{~b} =  \left( 1/\det{N} -1 \right) \epsilon^{bcd}
\frac{\varphi_c \hat{\cal D}_a\varphi_d}{\varphi^2}. 
\ee
The matter field covariant derivatives then have the form
\bea
\nabla_a S &=& {\cal D}_a S = N_a^{-1 b} \hat{\nabla}_b S \cr
\nabla_a \psi_i &=& N_a^{-1 b} \hat{\nabla}_b \psi_i ,
\eea
with the scalar and fermion fields' partially covariant derivatives defined
as
\bea
\hat{\nabla}_a S &=& \hat{\cal D}_a S \cr
\hat{\nabla}_a \psi_i &=& \hat{\cal D}_a \psi_i -\frac{i}{2}
\hat{\Gamma}_a^{~b} \gamma_{bij} \psi_j  .
\eea

The matter field action can be written in terms of the partially covariant
derivatives but the Nambu-Goto dreibein and metric are needed in order to
restore full $G$-invariance.  Using the Nambu-Goto dreibein $N$, its inverse
is
\be
N_a^{-1 b} = \delta_a^{~b} - \frac{1}{\cosh{\sqrt{v^2}} + d_c v^c}
\left(N_a^{~b} - \delta_a^{~b}\right) ,
\ee
where
\be
d_a = \varphi_a \frac{\sinh{\sqrt{v^2}}}{\sqrt{v^2}}
\ee
and $N$ is given in equation (\ref{ngdreibein2})
\be
N_a^{~b}= P_{vTa}^{~~~~b} + \cosh{\sqrt{v^2}} P_{vLa}^{~~~~b} + d_a v^b .
\ee
Upon application of the $\omega_Z = 0$ $G$-covariant constraint, these
dreibeine reduce to
\bea
N_a^{~b} &=& \delta_a^{~b} + \frac{1-\cosh{\sqrt{v^2}}}{\cosh{\sqrt{v^2}}}
\frac{v_a v^b}{v^2} \cr
N_a^{-1 b} &=& \delta_a^{~b} + \left( \cosh{\sqrt{v^2}} -1 \right)
P_{vLa}^{~~~~b} .
\eea
This can be used to form the Nambu-Goto metric
\bea
n^{ab} &=& N_c^{-1 a} \eta^{cd} N_d^{-1 b} \cr
&=& \eta^{ab} + \frac{v^a v^b}{v^2} \sinh^2{\sqrt{v^2}} .
\eea
The invariant interval can be written as
\be
ds^2 = dx^m g_{mn} dx^n = \omega^a \eta_{ab} \omega^b = \hat{\omega}^a
n_{ab} \hat{\omega}^b ,
\ee
with $n_{ab} = N_a^{~c} \eta_{cd} N_b^{~d} = \eta_{ab} - \frac{v_a v_b}{v^2}
\tanh^2{\sqrt{v^2}}$.

The fully $G$-invariant kinetic energy term for the scalar field in terms of
the partially covariant derivatives then becomes
\be
\nabla_a S \eta^{ab} \nabla_b S = \hat{\nabla}_a S n^{ab} \hat{\nabla}_b S .
\ee
Likewise, the fully $G$-invariant fermion kinetic energy becomes
\be
i\bar\psi \gamma^a \nabla_a \psi = i\bar\psi \hat{\gamma}^a \hat{\nabla}_a
\psi ,
\ee
where the Dirac matrices in the partially covariant basis are defined by
means of the Nambu-Goto dreibein
\be
\hat{\gamma}^a = \gamma^b N_b^{-1 a} .
\ee
The fully $G$-invariant matter action then takes the form
\be
\Gamma_{\rm matter} = \Gamma_S + \Gamma_f = \int d^3 x \det{e}~~{\cal
L}_{\rm matter} ,
\ee
where the invariant matter lagrangian can be written as
\bea
{\cal L}_{\rm matter} &=& {\cal L}_S + {\cal L}_f \cr
&=& \hat{\nabla}_a S n^{ab} \hat{\nabla}_b S -V(S) \cr
& &\qquad\quad +i\bar\psi \hat{\gamma}^a \hat{\nabla}_a \psi - m \bar\psi
\psi + Y(S, \bar\psi \psi) .
\eea

\bigskip
~\\
\section*{Acknowledgements}
\noindent
MN would like to thank Koji Hashimoto and
TtV would like to thank Mikhail Shifman, Arkady Vainshtein and
Mikhail Voloshin for interesting discussions.  The work of TEC
and MN was supported in part by the U.S. Department of Energy
under grant DE-FG02-91ER40681 (Task B).

\pagebreak

\setcounter{newapp}{1}
\setcounter{equation}{0}
\renewcommand{\theequation}{\thenewapp.\arabic{equation}}

\section*{\large Appendix A: \,  Conventions}

The D=3 Dirac matrix conventions are (labelling all space-time indices by
$m, n, \ldots = 0, 1, 2$ for convenience here):
\bea
\eta_{mn} &=& (+,-,-) \qquad   ; \qquad  \epsilon^{012} =+1  \cr
\epsilon_{mnl}\epsilon^{lrs} &=& \delta_m^r \delta_n^s -\delta_m^s
\delta_n^r \cr
\epsilon_{mnl}\epsilon^{nls} &=& +2\delta_m^s \cr
\gamma^m &=& (\sigma^2 , i\sigma^1 , i\sigma^3 )= -\gamma^{m\star}  \qquad ; 
\qquad \sigma^1
\sigma^2 = i \sigma^3  \cr
C &=& \gamma^0 = \sigma^2 =C^{-1}   \qquad  ;  \qquad C^T = -C  \cr
\gamma^m \gamma^n &=& \eta^{mn} 1 +i\epsilon^{mnr} \gamma_r \cr
\gamma^m \gamma^n \gamma^r &=& + i \epsilon^{mnr} 1 +\eta^{mn} \gamma^r
-\eta^{mr} \gamma^n +\eta^{nr} \gamma^m \cr
\{\gamma^m ,  \gamma^n \} &=& +2 \eta^{mn}  \cr
\gamma^{mn} &\equiv & \frac{i}{2} [ \gamma^m , \gamma^n ] =-\epsilon^{mnr}
\gamma_r \cr
\gamma^{mT} &=& -\gamma^0 \gamma^m \gamma^0 = -\gamma^{m\dagger} \qquad ; 
\qquad (\gamma^0
\gamma^m)^T = \gamma^0 \gamma^m  \cr
\gamma_{ij}^{m} \gamma_{ji}^n & = & 2 \eta^{mn} \nonumber \cr
\gamma_{ij}^{k} \gamma_{jk}^l \gamma_{ki}^m & = & 2 i \epsilon^{klm} \cr
\gamma^{12} &=& -\sigma^2 = -\gamma^{21}  \cr
\gamma^{01} &=& i\sigma^3 = -\gamma^{10}  \cr
\gamma^{02} &=& - i\sigma^1 = -\gamma^{20}  \cr
\gamma^m C &=& ( 1, - \sigma^3 , \sigma^1 ) \cr
(\gamma^m C)_{ij}A_m &=& \pmatrix{
(A_0 - A_1) & A_2 \cr
A_2 & (A_0 + A_1) }_{ij}  .
\eea
\be
~~
\ee

Conventions involving the two-component real Grassmann variable (Majorana) 
fields
$\theta_i$ and $\lambda_i$ are given below in terms of $\theta$ with
corresponding formulae for all such anti-commuting variables.  The
derivative with respect to
$\theta_i$ is defined through the Taylor expansion formula
\be
f(\theta +\delta\theta ) = f(\theta) +\delta\theta_i
\frac{\partial}{\partial
\theta_i} f(\theta) .
\ee
Hence the derivative is given by
\be
\frac{\partial}{\partial \theta_i} \theta_j = \delta_{ij} .
\ee
In a similar manner the conjugate Majorana spinor $\bar\theta_i$ is defined 
as
\be
\bar\theta_i \equiv (\gamma^0 )_{ij} \theta_j = C_{ij} \theta_j .
\ee
(For complex spinors (matter fields), $\psi_i$, the adjoint is defined as 
usual
$\bar\psi = \psi^\dagger \gamma^0$.)  The derivative with
respect to $\bar\theta$ is defined analogously
\be
\frac{\partial}{\partial \bar\theta_i} \bar\theta_j = \delta_{ij} .
\ee
In other words
\be
\frac{\partial}{\partial \bar\theta_i} = - (\gamma^0 )_{ij}
\frac{\partial}{\partial \theta_j} ,
\ee
so that the mixed derivative formulas are
\bea
\frac{\partial}{\partial \bar\theta_i} \theta_j &=& - (\gamma^0 )_{ij} \cr
\frac{\partial}{\partial \theta_i} \bar\theta_j &=& - (\gamma^0 )_{ij} .
\eea
Consequently the product of spinors
\be
\left( \bar\theta_i \psi_i \right)^\dagger = \bar\theta_i \psi_i .
\ee
Further
\be
\theta_i \theta_j =\frac{1}{2} \bar\theta \theta \gamma^0_{ij} ,
\ee
which leads to
\bea
\bar\theta \gamma^m \theta &=& 0 \cr
\bar\theta \gamma^m \gamma^n \theta &=& \bar\theta \theta \eta^{mn} \cr
\bar\theta \gamma^m \gamma^n \gamma^r \theta &=& i \bar\theta\theta
\epsilon^{mnr} .
\eea

Projectors can be defined in terms of the D=3 vector $v^m$ as
\bea
P_{vT}^{mn} &=& \left( \eta^{mn} - \frac{v^m v^n}{v^2}\right) \cr
P_{vL}^{mn} &=& \frac{v^m v^n}{v^2} .
\eea

\newpage

\newpage

\begin{thebibliography}{99}

\bibitem{Coleman:sm}
S.~R.~Coleman, J.~Wess and B.~Zumino,
Phys.\ Rev.\  {\bf 177}, 2239 (1969);
C.~G.~Callan, S.~R.~Coleman, J.~Wess and B.~Zumino,
Phys.\ Rev.\  {\bf 177}, 2247 (1969).

\bibitem{Volkov73}
D.~V.~Volkov, Sov.\ J.\ Particles and Nuclei {\bf 4}, 3
(1973); V.~I.~Ogievetsky, in
{\sl Recent developments in relativistic quantum field theory
and its applications}, edited by J.~Lukierski,
proceedings of the X-th Winter School of
Theoretical Physics in Karpacz (Universitas Wratislaviensis,
Wroclaw, 1974) Vol.~1, p. 227.

\bibitem{Ivanov:1975zq}
E.~A.~Ivanov and V.~I.~Ogievetsky,
Teor.\ Mat.\ Fiz.\  {\bf 25}, 164 (1975).

\bibitem{Dirac:1962iy}
P.~A.~Dirac,
Proc.\ Roy.\ Soc.\ Lond.\ A {\bf 268}, 57 (1962).

\bibitem{Nambu:1974zg}
Y.~Nambu,
Phys.\ Rev.\ D {\bf 10}, 4262 (1974).

\bibitem{Goto:ce}
T.~Goto,
Prog.\ Theor.\ Phys.\  {\bf 46}, 1560 (1971).

\bibitem{Born:gh}
M.~Born and L.~Infeld,
Proc.\ Roy.\ Soc.\ Lond.\ A {\bf 144}, 425 (1934).

\bibitem{West:2000hr}
P.~C.~West,
JHEP {\bf 0002}, 024 (2000)
[arXiv:hep-th/0001216].

\bibitem{Hughes:fa}
J.~Hughes, J.~Liu and J.~Polchinski,
Phys.\ Lett.\ B {\bf 180}, 370 (1986).

\bibitem{Hughes:dn}
J.~Hughes and J.~Polchinski,
Nucl.\ Phys.\ B {\bf 278}, 147 (1986).

\bibitem{Duff:1996zn}
M.~J.~Duff,
Published in ``Fields, Strings an Duality,'' TASI'96
Proceedings, World Scientific, (1997)
[arXiv:hep-th/9611203].

\bibitem{West:1998ey}
P.~C.~West,
arXiv:hep-th/9811101.

\bibitem{Bagger:1994vj}
J.~Bagger and A.~Galperin,
Phys.\ Lett.\ B {\bf 336}, 25 (1994)
[arXiv:hep-th/9406217];
J.~Bagger and A.~Galperin,
Phys.\ Rev.\ D {\bf 55}, 1091 (1997)
[arXiv:hep-th/9608177];
J.~Bagger and A.~Galperin,
arXiv:hep-th/9810109.

\bibitem{deAzcarraga:gm}
J.~A.~de Azcarraga, J.~P.~Gauntlett, J.~M.~Izquierdo and P.~K.~Townsend,
Phys.\ Rev.\ Lett.\  {\bf 63}, 2443 (1989);
E.~R.~Abraham and P.~K.~Townsend,
Nucl.\ Phys.\ B {\bf 351}, 313 (1991);
M.~Cvetic, F.~Quevedo and S.~J.~Rey,
Phys.\ Rev.\ Lett.\  {\bf 67}, 1836 (1991);
G.~R.~Dvali and M.~A.~Shifman,
Phys.\ Lett.\ B {\bf 396}, 64 (1997)
[Erratum-ibid.\ B {\bf 407}, 452 (1997)]
[arXiv:hep-th/9612128];
A.~Kovner, M.~A.~Shifman and A.~Smilga,
Phys.\ Rev.\ D {\bf 56}, 7978 (1997)
[arXiv:hep-th/9706089];
A.~Ritz, M.~Shifman and A.~Vainshtein,
arXiv:hep-th/0205083;
D.~Binosi and T.~ter Veldhuis,
Phys.\ Lett.\ B {\bf 476}, 124 (2000)
[arXiv:hep-th/9912081].
M.~Naganuma, M.~Nitta and N.~Sakai,
Phys.\ Rev.\ D {\bf 65}, 045016 (2002)
[arXiv:hep-th/0108179];
M.~Naganuma and M.~Nitta,
Prog.\ Theor.\ Phys.\  {\bf 105}, 501 (2001)
[arXiv:hep-th/0007184].

\bibitem{Ivanov:1999fw}
E.~Ivanov and S.~Krivonos,
Phys.\ Lett.\ B {\bf 453} (1999) 237
[arXiv:hep-th/9901003];
S.~Bellucci, E.~Ivanov and S.~Krivonos,
Phys.\ Lett.\ B {\bf 482}, 233 (2000)
[arXiv:hep-th/0003273];
S.~Bellucci, E.~Ivanov and S.~Krivonos,
Nucl.\ Phys.\ Proc.\ Suppl.\  {\bf 102}, 26 (2001)
[arXiv:hep-th/0103136].

\bibitem{Ivanov:2001gd}
E.~Ivanov,
Theor.\ Math.\ Phys.\  {\bf 129}, 1543 (2001)
[Teor.\ Mat.\ Fiz.\  {\bf 129}, 278 (2001)]
[arXiv:hep-th/0105210];
E.~Ivanov,
in {\sl New Development in Fundamental Interaction Theories},
edited by J.~Lukierski and J. Rembielinski,
AIP Conf.~Proc.~{\bf 589} (AIP, Melville, NY, 2001), p.~61;
E.~Ivanov,in {\sl New Symmetries and Integrable Models},
edited by A.~Frydryszak, J.~Lukierski, Z.~Popowicz 
(World Scientific, Singapore, 2000) p. 246
[arXiv:hep-th/0002204].

\bibitem{Sorokin:1999jx}
See for example
D.~P.~Sorokin,
Phys.\ Rept.\  {\bf 329}, 1 (2000)
[arXiv:hep-th/9906142], and references therein.

\bibitem{Chibisov:1997rc}
B.~Chibisov and M.~A.~Shifman,
Phys.\ Rev.\ D {\bf 56}, 7990 (1997)
[Erratum-ibid.\ D {\bf 58}, 109901 (1998)]
[arXiv:hep-th/9706141].

\bibitem{Sakamura:2002iw}
Y.~Sakamura,
arXiv:hep-th/0207159.

\bibitem{Volkov:jx}
D.~V.~Volkov and V.~P.~Akulov,
JETP Lett.\  {\bf 16}, 438 (1972).

\bibitem{Duff:1992hu}
M.~J.~Duff and J.~X.~Lu,
Nucl.\ Phys.\ B {\bf 390}, 276 (1993)
[arXiv:hep-th/9207060];
P.~K.~Townsend,
Phys.\ Lett.\ B {\bf 373}, 68 (1996)
[arXiv:hep-th/9512062];
C.~Schmidhuber,
Nucl.\ Phys.\ B {\bf 467}, 146 (1996)
[arXiv:hep-th/9601003];

\bibitem{Aganagic:1996nn}
M.~Aganagic, C.~Popescu and J.~H.~Schwarz,
Nucl.\ Phys.\ B {\bf 495}, 99 (1997)
[arXiv:hep-th/9612080].

\bibitem{Hou:1999rv}
X.~R.~Hou, A.~Losev and M.~A.~Shifman,
(domain
walls in theories with compactified dimensions),''
Phys.\ Rev.\ D {\bf 61}, 085005 (2000)
[arXiv:hep-th/9910071];
R.~Hofmann and T.~ter Veldhuis,
Phys.\ Rev.\ D {\bf 63}, 025017 (2001)
[arXiv:hep-th/0006077];
N.~Maru, N.~Sakai, Y.~Sakamura and R.~Sugisaka,
Phys.\ Lett.\ B {\bf 496}, 98 (2000)
[arXiv:hep-th/0009023];
N.~Maru, N.~Sakai, Y.~Sakamura and R.~Sugisaka,
Nucl.\ Phys.\ B {\bf 616}, 47 (2001)
[arXiv:hep-th/0107204];
N.~Sakai and R.~Sugisaka,
models,''
Phys.\ Rev.\ D {\bf 66}, 045010 (2002)
[arXiv:hep-th/0203142].

\bibitem{Clark:1996aw}
T.~E.~Clark and S.~T.~Love,
Phys.\ Rev.\ D {\bf 54}, 5723 (1996)
[arXiv:hep-ph/9608243];
T.~E.~Clark and S.~T.~Love,
Phys.\ Rev.\ D {\bf 63}, 065012 (2001)
[arXiv:hep-th/0007225].
\end{thebibliography}
\end{document}